\documentclass{elsart1p}

\usepackage{graphicx}

\usepackage{amssymb}

\begin{document}

\begin{frontmatter}



\title{Progress and Challenges in the Theory of Nuclei\thanksref{label1}}
\thanks[label1]{Oak Ridge
National Laboratory is managed by UT-Battelle for the U.S. Department
of Energy under Contract No.~DE-AC05-00OR22725.}


\author{D.J. Dean}

\address{Physics Division, Oak Ridge National Laboratory, P.O. Box 2008, 
Oak Ridge, TN 37831-6373 USA}

\begin{abstract}
Nuclear theory today aims for a comprehensive theoretical framework that
can describe all nuclei. I discuss recent progress in this pursuit and 
the associated challenges as we move forward. 
\end{abstract}

\end{frontmatter}

\section{The journey toward a comprehensive theory of nuclei}

Nuclei comprise 99.9\% of all baryonic matter in the Universe 
and are the fuel that burns in stars \cite{schiffer}. The rather 
complex nature of the nuclear forces among protons 
and neutrons generates a broad range and diversity 
in the nuclear phenomena that we observe. Experiments
indicate that developing a 
comprehensive description of all nuclei and their 
reactions requires theoretical and experimental 
investigations of rare isotopes with unusual 
neutron-to-proton ratios that are very different from their
stable counterparts.  These rare nuclei are difficult to 
produce and study experimentally since they can have 
extremely short lifetimes. 
The goal of a comprehensive description and reliable 
modeling of nuclei represents one of the great 
intellectual opportunities for nuclear physics in the coming years. 

Key scientific themes being addressed by the physics of nuclei
and nuclear astrophysics communities are captured by five 
overarching questions:
\begin{itemize}
\item What is the nature of the nuclear force that 
binds protons and neutrons into stable nuclei and rare isotopes?
\item What is the origin of simple patterns in complex nuclei?
\item What is the nature of neutron stars and dense nuclear matter?
\item What is the origin of the elements in the cosmos?
\item What are the nuclear reactions that drive stars and stellar explosions?
\end{itemize}
These questions align well with the drivers of rare isotope 
science as recently summarized in a report of the U. S. 
National Academies of Science Rare Isotope Scientific Assessment Committee 
(RISAC) \cite{risac}.  One primary aspect of the first and second 
question concerns testing the predictive power of 
models by extending experiments to new regions of mass 
and proton-to-neutron ratio and identifying new 
phenomena that will challenge existing many-body theory. 

One of those challenges concerns the concept of nuclear shell structure and 
magic numbers. Understanding the origin of the classic, stable, 
nuclear magic numbers  earned Goeppert-Mayer and Jensen the
1963 Nobel Prize. At the time, there were no data on 
very neutron-rich nuclei; however, theoretical predictions and 
experimental discoveries in the last decade indicate that 
nuclear shell structure is a rather fluid concept.  For example, 
experimental data indicate that the magic numbers at N=20 and 28 fade away with 
neutron number, and the new magic numbers at N=14, 16, and 32 seem 
to appear (see, for example Ref. \cite{fridmann05}). Nuclei far from 
stability also exhibit unusual properties as compared to 
their stable cousins.  For example, the radial extension 
of the two-neutron halo in $^{11}$Li (where $^9$Li acts as a core)
is the same as that of $^{208}$Pb. 

We can conclude from just these two examples that nuclear structure 
is changing in the exotic environment.  Since nuclei are self-bound,
they generate their own mean field which 
determines how shell orbitals are filled. The experimental evidence indicates
the way orbits are being filled is changing in neutron-rich nuclei. 
Furthermore, many-body correlations such as pairing superfluidity become 
crucial when the binding energy becomes small. These correlations 
also affect the open-shell character of nuclei, since the 
continuum of scattering states lies very close to the bound
state.  

One of the interesting features of nuclear physics today is the 
overlaps and synergy between the physics of nuclei and nuclear 
astrophysics. The changes in nuclear structure occurring away from
the valley of beta-stability do affect nucleosynthesis pathways. 
We have initial evidence of this on the measurements of the 
lifetime of $^{78}$Ni \cite{ni78} and its impact on 
r-process nucleosynthesis. Other examples involve the
impact of electron capture on the core collapse mechanism of 
type-II supernovae \cite{langanke03} and the challenge of
obtaining a nuclear equation of state that can quantitatively 
be used in predictions of neutron star properties \cite{muther07}. 

A robust theoretical capability is required if we 
want to understand stable and exotic nuclei that are 
the core of matter and the fuel of stars. I will devote
the remainder of this Proceedings to a discussion of some
of the issues that remain in building this capability. 

\section{Progress and challenges in the theory of nuclei}

The nuclear many-body problem spans nuclei from A=2 (the deuteron) to the
superheavy region, and different theoretical techniques are pursued in
different regions of the chart of nuclei. This is depicted in Fig. 1
where the current reach of various theoretical methods is interposed in 
the chart of nuclei. The challenges for theory can be viewed
broadly as: determining the nuclear interaction, calculating ab initio 
nuclei, including the continuum, and finding an appropriate energy 
density functional for a nuclear Density Functional Theory approach. 
I will briefly discuss these areas in the following subsections, 
drawing from some of the work in which I have been involved. 

\begin{figure}[t]
\begin{center}
\includegraphics[width=0.6\textwidth,angle=0]{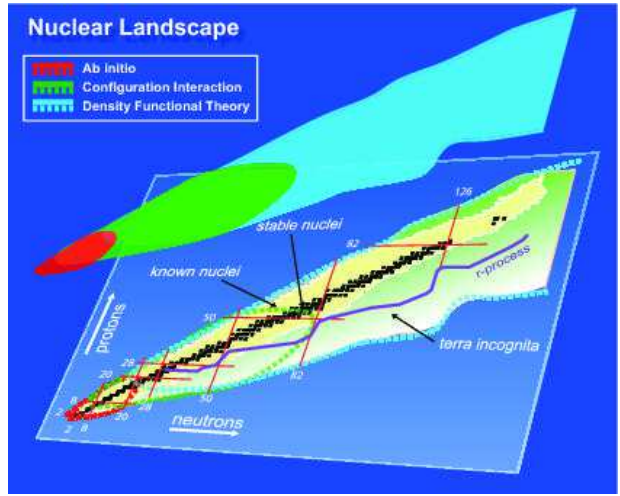}
\caption{(Color online) 
The theoretical methods and computational techniques used to solve the 
nuclear many-body problem. 
The red vertical and horizontal lines show the magic 
numbers, reflecting regions where nuclei are expected to be 
more tightly bound and have longer half-lives. The 
anticipated path of the astrophysical r-process responsible 
for nucleosynthesis of heavy elements is also shown (purple line). 
The thick dotted lines indicate domains of major 
theoretical approaches to the nuclear many-body problem. 
For the lightest nuclei, ab initio calculations (Green's function 
Monte Carlo, no-core shell model, coupled cluster method), based 
on the bare nucleon-nucleon interaction, are possible (red). 
Medium-mass nuclei can be treated by configuration 
interaction techniques [interacting shell model (green)]. 
For heavy nuclei, the density functional theory based on 
self-consistent/mean-field theory (blue) is the tool of choice. 
By investigating the intersections between these theoretical 
strategies, one aims to develop a unified description of the nucleus.}
\end{center}
\label{fig1}
\end{figure}

\subsection{Determining nuclear interactions}

The forces among nucleons determine how nuclei behave. These 
forces arise from QCD. While there is some effort to extract
directly from lattice calculations \cite{ishii07} the nuclear scattering, there
have also been important developments in the application of a chiral Effective
Field Theory (EFT) to the nuclear problem 
\cite{weinberg,bira}. An EFT was developed
for nuclei from a chiral Lagrangian that maintains all the symmetries of 
QCD and treats as fundamental particles the pions and nucleons. A 
power-counting scheme was developed so that the expansion could
be carried out in a systematic way using as a small 
parameter of expansion the momentum transfer in nucleon-nucleon 
scattering divided by the QCD energy scale (1 GeV).  This expansion
has been carried out to fourth-order (known as N$^3$LO) \cite{n3lo}. The 
resulting nucleon-nucleon potential fits all scattering phase
shifts up to the cut-off used in the EFT. 

An interesting feature of the EFT is that three-nucleon interactions
arise naturally at the third order (N$^2$LO). Chiral three-body 
forces have been known for a long time, but their systematic 
computation was not known until recently. Nuclear spectroscopy using
three-body interactions from the EFT expansion improve when compared
to calculations with two-body forces alone \cite{navratil07}. This is also 
to be expected since it has long been known that 
two-body forces alone do not reproduce the correct binding in $^4$He. 

Thus, our starting point is a Hamiltonian of the form
\begin{equation}
H=t-t_{cm} + V_2 + V_3 \;,
\end{equation}
where $t$ is the kinetic energy operator and $t_{cm}$ is the 
center-of-mass operator (so that the kinetic energy maintains 
translational invariance)
and the $V_2$ and $V_3$ represent the two-body and three-body 
nuclear potentials.  The three-body interaction 
can actually be written in normal-ordered form as
\begin{equation}
\label{normal}
V_3 = {1\over 6} \sum_{ijk} \langle ijk||ijk\rangle
+ {1\over 2} \sum_{ijpq} \langle ijp||ijq\rangle
\{\hat{a}^\dagger_p \hat{a}_q \} \\\nonumber
 + {1\over 4} \sum_{ipqrs} \langle ipq||irs\rangle
\{\hat{a}^\dagger_p \hat{a}^\dagger_q \hat{a}_s \hat{a}_r \} + \hat{h}_3 \,,
\end{equation}
where $\langle \alpha\beta\gamma||\delta\epsilon\eta\rangle$ represents
a matrix element of the three-body interaction, $i,j,k$ denote
the hole states, $a,b,c$ denote the particle states, 
$p,q,r$ denote all states,
and $\hat{h}_3$ denotes the residual three-body Hamiltonian
\begin{equation}
\label{h3}
\hat{h}_3\equiv {1\over 36} \sum_{pqrstu} \langle pqr||stu\rangle
\{\hat{a}^\dagger_p \hat{a}^\dagger_q \hat{a}^\dagger_r \hat{a}_u \hat{a}_t
\hat{a}_s \} \,.
\end{equation}
This notation will be useful in what follows. 

$V_2$ is often modeled with a very repulsive
short distance interaction. In the case of Av18 \cite{bob95}, the
central repulsion at $r=0$ is approximately 3 GeV; in 
N$^3$LO, it is 500 MeV \cite{n3lo}. In order to 
use these interactions efficiently in a framework where 
one expands the many-body wave function in a basis, 
it is desirable to renormalize these interactions. 
This can be done in several ways, including the $G$-matrix 
approach \cite{mhj1995} and
similarity transform methods \cite{leesuz}. 

More recently, work in
the direction of renormalization group (RG) \cite{bog03,bog07} techniques has
been used to evolve the nucleon-nucleon interactions from high momentum to 
low momentum while preserving the nucleon-nucleon phase shifts up a given
momentum.  These interactions can then be projected onto the 
appropriate basis states for computation. These approaches offer promise
with a price: the resulting two-body interactions do not recover 
the binding energy that would be obtained from a pure two-body 
calculation, and one must indeed include a three-body force in
the calculations. Furthermore, the RG interactions tend toward 
overbinding both in nuclei \cite{hagen_c} and 
in nuclear matter \cite{bogner05}. The cure to this problem is 
the introduction of three-body forces which naturally occur
in the RG approaches. Another possibility is to adjust the 
three-body force that comes from EFT to obtain the experimental binding of
$^4$He \cite{nogga}.

\subsection{Calculating {\em ab initio} Nuclei}

For light nuclei with mass numbers $A \le 12$, both Green's
function Monte Carlo methods \cite{pieper02} and 
no-core shell-model calculations \cite{navratil00,navratil07} 
using a basis-state expansion provide almost converged benchmarks 
for selected two- and three-body Hamiltonians. 
The agreement with experimental data for many 
light nuclei is quite reasonable
in these calculations. GFMC techniques do not need to 
utilize renormalized interactions, but they can only use 
local potentials, and the main group involved uses one two-body 
potential (Av18, \cite{bob95}) and an appropriate three-body 
potential fitted to experimental data in light nuclei \cite{il2}. 
The NCSM approach can use any potential, but in most cases
uses renormalized ones in order to obtain converging calculations. 
Both methods exhibit exponential scaling with the number of particles 
(to reach the same accuracy as in $^4$He with the same methods). 
Recently, we have seen a movement of these methods into the realm of
light nuclear reactions. For example, the GFMC approach was 
used to describe $n+\alpha$ scattering \cite{nollett07} while
the NCSM was used to calculate the $^7$Be($p,\gamma$)$^8$B S-factor
\cite{nav06}. Recent progress has also been made in ab initio 
inelastic four-body scattering problems \cite{deltuva07}.

Coupled-cluster theory \cite{coester,zab,cizek,bogdan} 
represents another approach 
to ab initio calculations of nuclei. Along with collaborators, 
including computational chemists, we have developed the theory
and methods~\cite{dhj,kow,hagen_a,hagen_b,hagen_c} to 
the point where we can calculate nuclear 
properties in medium-mass nuclei. Nuclear coupled-cluster theory exhibits
polynomial scaling with the size of the basis state and number of 
particles, and is size-extensive. This very important property 
means that no unlinked diagrams enter into the theory. 

Coupled-cluster theory starts with the simple assertion 
that the correlated ground-state wave function can be 
described by applying an exponentiated correlation operator
to an uncorrelated Slater determinant that naively describes the
nucleus: 
\begin{equation}
\mid\Psi\rangle = \exp\left(T\right)\mid\Phi\rangle \;. 
\end{equation}
The correlation operator $T$ induces various $n$p-$n$h (particle-hole) 
correlations (up to the number of nucleons, $A$, in the nucleus): 
\begin{equation}
T=T_1 + T_2 + \cdots T_A \;.
\end{equation}
The energy of the ground state is given by 
\begin{equation}
E=\langle\Phi\mid\exp(-T)H\exp(T)\mid\Phi\rangle = 
\langle\Phi\mid\bar{H}\mid\Phi\rangle = 
\langle\Phi\mid \left(H\exp(T)\right)_c\mid\Phi\rangle \;, 
\label{energy}
\end{equation}
where the subscript $c$ means that only connected diagrams enter
into the expressions.  The $T_n$ operators take the form (as is
the case here for the two-body excitation operator)
\begin{equation}
T_2 = \sum_{ab,ij}t^{ab}_{ij} a^\dagger_aa^\dagger_ba_j a_i
\end{equation}
where $t^{ab}_{ij}$ are the 
$2p$-$2h$ correlation amplitudes. 

One must compute the correlation amplitudes through
equations that left-project $\bar{H}$ onto the space of excited 
Slater determinants. For example, if the theory only contains
$T_1$ and $T_2$ operators, then the equations to solve for the 
coefficients $t^a_i$ and $t^{ab}_{ij}$ are 
\begin{eqnarray}
0&=&\langle\Phi^a_i\mid\bar{H}\mid\Phi\rangle \nonumber \\
0&=&\langle\Phi^{ab}_{ij}\mid\bar{H}\mid\Phi\rangle\;, 
\end{eqnarray}
and so on. 

Eqn.~(\ref{energy}) is exact; however, 
the power of the coupled-cluster theory is its highly 
accurate computation of the energy even when one limits to 
lower order the number of $T_n$ operators in the theory. 
For example, at the $T_1$ and $T_2$ levels (coupled clusters
in singles and doubles, or CCSD), approximately 90\% of the
correlation is obtained, while with approximate triples corrections
(denoted CCSD(T)), nearly all of the correlation
energy is obtained \cite{kow,hagen_c}. This contrasts sharply
to truncated shell-model calculations where the truncation introduces
unlinked diagrams causing growing errors with particle number
\cite{rod07}.

A certain amount of benchmarking of methods is necessary in order
to understand their regions of validity. For coupled-cluster theory, 
this has recently been performed in the triton and $^4$He. We also 
performed extremely large basis-set calculations to which other 
methods can compare when they reach the technical capability to 
do so \cite{hagen_c}.  We summarize our coupled-cluster results 
for the binding energies of $^4$He, $^{16}$O, and $^{40}$Ca 
in Table~\ref{tab:corr_energy},
which gives the extrapolated correlation energies $\Delta E_{\rm CCSD}$
and $\Delta E_{\rm CCSD(T)}$, relative to the uncorrelated energy 
$E_0=\langle\Phi\mid H\mid\Phi\rangle$. We find that for $^4$He, $^{16}$O, and
$^{40}$Ca, the triples corrections are a factor of $\approx 0.015,
0.066$, and $0.081$ smaller than the CCSD correlation energies. From
this, we estimate the missing correlation energy from quadruples,
pentuplets, and so on  to be of the order of $1$~MeV for $^{40}$Ca.
We note that $^{16}$O is overbound by about $20$~MeV, and $^{40}$Ca
by about $150$~MeV  when compared to the experimental binding energies.
This is not surprising and points to the importance of 3NF for
nuclear structure calculations and to its apparently large role 
in the RG prescriptions.

\begin{table}[t]
\begin{center}
\begin{tabular}{l|r|r|r}
\hline
\multicolumn{1}{c|}{} & \multicolumn{1}{c|}{$^4$He} &
\multicolumn{1}{c|}{$^{16}$O} & \multicolumn{1}{c}{$^{40}$Ca} \\
\hline \hline
$E_0$ & -11.8 & -60.2 & -347.5 \\
$\Delta E_{\rm CCSD}$ & -17.1 & -82.6 & -143.7 \\
$\Delta E_{\rm CCSD(T)}$ & -0.3 & -5.4 & -11.7 \\
\hline
$E_{\rm CCSD(T)}$ & -29.2 & -148.2 & -502.9 \\
\hline \hline
exact (FY) & -29.19(5) & & \\
\hline
\end{tabular}
\caption{Reference vacuum energies, $E_0$, CCSD and CCSD(T) correlation
energies, $\Delta E_{\rm CCSD}$ and $\Delta E_{\rm CCSD(T)}$, and binding
energies $E_{\rm CCSD(T)}$ for $^4$He, $^{16}$O and $^{40}$Ca. The vacuum
energies, $E_0$,
are for $\hbar\omega = 14$~MeV in the case of $^4$He and $\hbar\omega
= 22$~MeV for $^{16}$O and $^{40}$Ca.
The CCSD and CCSD(T) energies are the extrapolated infinite model space
results. The exact Faddeev-Yakubovsky result is from Ref.~\cite{nogga}}
\label{tab:corr_energy}
\end{center}
\end{table}

The three-body force is a key component of modern nuclear Hamiltonians. 
Other methods have been adapted to include three-body interactions already
several years ago.  We recently developed coupled-cluster 
theory for three-body forces \cite{hagen_b}. This represents 
a major improvement to the scientific reach of modern 
coupled-cluster calculations for nuclei.
An exponential extrapolation of the (approximate) CCSD(T) minima to an
infinite model space is shown in Fig. 2 and yields $E_\infty
= -28.24$~MeV. This is in excellent agreement with the exact
Faddeev-Yakubovsky result $E=-28.20(5)$~MeV. In our largest model
space at the minimum $\hbar\omega = 17$~MeV, the ground-state
expectation values for the center-of-mass Hamiltonian is $\langle
H_{\rm cm} \rangle \approx 20$~keV while the expectation value for the
angular momentum is zero for a closed-shell nucleus by construction.

\begin{figure}[t]
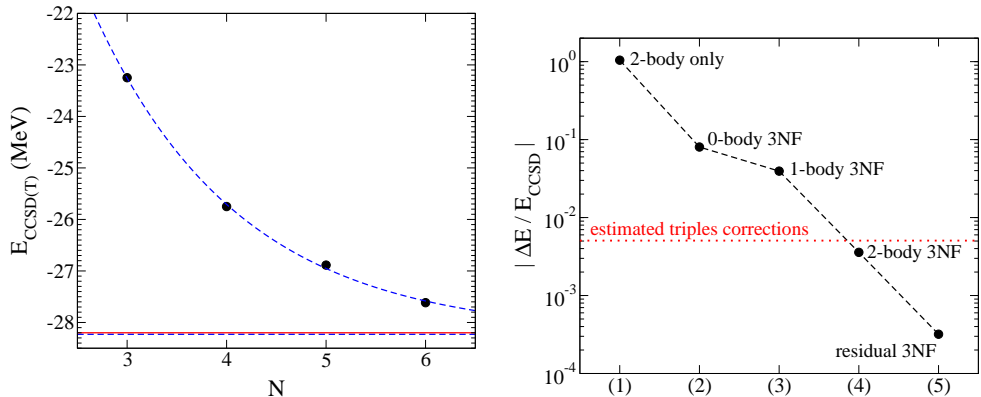

\includegraphics[width=0.46\textwidth,angle=0]{figure2a.eps}
\hspace{0.1in}
\includegraphics[width=0.46\textwidth,angle=0]{figure2b.eps}
\caption{Left: (Color online) Data points: CCSD(T) results
(taken at the $\hbar \omega$ minima) for the binding energy
of $^4$He with 3NFs as a function of the
number of oscillator shells. Dashed lines: Exponential fit to
the data and asymptote of the fit. Full line: Exact result.
Right: Relative contributions $|\Delta E / E|$ to
the binding energy of $^4$He at the CCSD level. The different
points denote the contributions from
(1) low-momentum NN interactions,
(2) the vacuum expectation value of the 3NF,
(3) the normal-ordered one-body Hamiltonian due to the 3NF,
(4) the normal-ordered two-body Hamiltonian due to the 3NF,
and (5) the residual 3NFs. The dotted line estimates the
corrections due to omitted three-particle--three-hole clusters.}
\label{fig7b}
\end{figure}

It is interesting to analyze the different contributions $\Delta E$ to
the binding energy $E$. The individual contributions are given in
Fig. 2 for a model space of $N=4$ oscillator shells and
$\hbar \omega = 20$~MeV. The main contribution stems from the
low-momentum NN interaction.  The contributions from 3NFs account only
for about 10\% of the total binding energy. This is consistent with
the chiral EFT power-counting estimate $\langle V_{\rm 3N} \rangle
\sim (Q/\Lambda_\chi)^3 \: \langle V_{{\rm lowk\,}k} \rangle \approx 0.1 \:
\langle V_{{\rm low}\,k} \rangle$ \cite{nogga}. 
The second-, third-, and fourth-largest
contribution are due to the first, second, and third term on the
right-hand side of Eq.~(\ref{normal}). These are the density-dependent
zero-, one-, and two-body terms, which resulted from the normal
ordering of the three-body Hamiltonian in coupled-cluster theory.  The
contributions from the residual three-body Hamiltonian,
Eq.~(\ref{h3}), are very small and are represented by the last point
in the right panel of Fig. 2.  The residual 3NF contributes to the
energy directly and indirectly through a
modification of the $T_1$ and $T_2$ cluster amplitudes.  
Apparently, both contributions are very small. This is a promising
result, and time will tell whether it holds for heavier nuclei. 

\subsection{Presence of the continuum}

Exotic phenomena emerge in weakly bound and
resonant many-body quantum systems. These phenomena
include ground states that are embedded
in the continuum, melting and reorganizing of shell
structures, extreme matter clusters and halo
densities. 
A theoretical description of weakly bound and
unbound quantum many-body systems represents a challenging
undertaking.  The weakly bound character of these
systems means that they should be treated 
as open quantum systems where coupling
with the scattering continuum can take place. Furthermore, 
theoretical treatments that consider continuum basis states
can describe resonant widths (lifetimes). 

Recent work with Gamow states employed in Hamiltonian
diagonalization methods \cite{michel,betan,hagen06} 
have shown that these basis states correctly depict properties associated
with open quantum systems. This Berggren
basis is composed of bound, resonant, and (continuum)
scattering single-particle states \cite{berggren}. This basis
significantly improves and facilitates the description
of loosely bound systems and is essential in the
description of unbound systems. 
In addition, several
groups have worked on alternative methods, such
as the continuum shell model \cite{volya,rotter} and
the recently developed shell model embedded in the
continuum \cite{okolowicz}. 

We recently developed a complex version 
of the coupled-cluster method that 
can utilize the Gamow-basis to calculate 
widths of states in the He isotopic chain in 
an ab initio framework \cite{hagen_a}. We utilized
a V$_{{\rm low} k}$ interaction from N$^3$LO at a cut-off 
of 1.9~fm$^{-1}$ in a Berggren basis to perform 
the calculations. These novel calculations, 
the results of which are shown in Fig. 3,
while by no means perfect, utilizing only a two-body 
interaction, do indicate the power of moving 
beyond the shell model. 

\begin{figure}[t]
\begin{center}
\includegraphics[width=0.5\textwidth,angle=0]{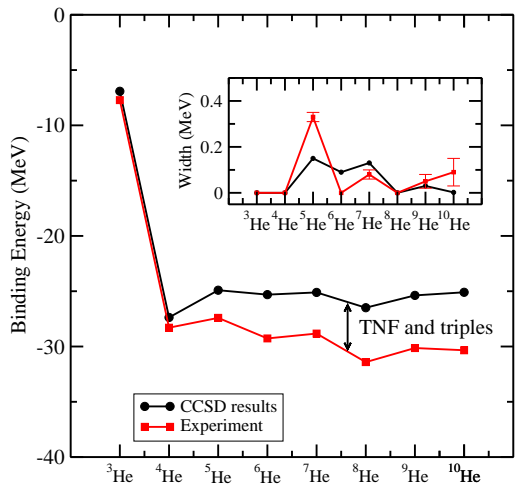}
\caption{(Color online) CCSD calculation of the $^{3-10}$He ground 
states with the low-momentum N$^3$LO nucleon-nucleon interaction 
for increasing number of partial waves. The energies $E$ are given in 
MeV for both real and imaginary parts. Experimental data are from
Ref. \cite{audi03}.  Our calculated width 
of $^{10}$He is $\approx$ 0.002~MeV.}
\end{center}
\label{fig3}
\end{figure}

\subsection{Density functional theory for nuclei}

Nuclear Density Functional Theory (nDFT) is being used and 
improved to describe heavier nuclei. The basic idea is to expand
the energy density functional to second order (gradient terms) in
the density and to calculate with this functional all nuclear 
properties.  One must determine the coupling constants
that characterize the functional from experimental data. Odd-$A$ and
odd-odd nuclei will be key to any new parameterizations as they 
allow one to interrogate time-odd terms in the energy density 
functionals. 

The nuclear DFT efforts have been quite successful 
in describing a wide variety of nuclear data with very 
good precision across the nuclear chart. The various parameterizations 
usually work quite well in regions where nuclear masses and 
other properties are experimentally determined, but 
extrapolations into very neutron-rich nuclei 
have been problematic. The next-generation experimental facilities should 
enable theorists to obtain a functional parameterization that 
will describe bulk properties of all nuclei. Various nuclear data 
along long isotopic and isotonic chains are needed to 
constrain the isovector part of the energy functional. 
More specifically, one needs (difference of) masses 
and measures of collectivity and of the shell evolution in unknown 
regions, where predictions of currently used functionals 
disagree. Initial attempts are represented in Fig. 4 where
the two-neutron separation energies were calculated from an
even-even mass table for the Sly4 Skyrme interaction with 
pairing included and partial number projection \cite{stoitsov03}. 
Other groups are also investigating nuclear mass systematics in
large mass-table calculations and have recently extended these calculations
to include a study of fission barriers \cite{goriely05}. 
Other data that could be included in 
determining the coupling constants of the energy density functional, such as
large deformations (at low and high angular momentum) 
and multipole strength distributions in 
neutron-rich nuclei, will also be extremely valuable. The quest for 
a universal energy density function represents one of the key challenges
of a U.S. Department of Energy SciDAC project. 

\begin{figure}[t]
\begin{center}
\includegraphics[width=0.5\textwidth,angle=0]{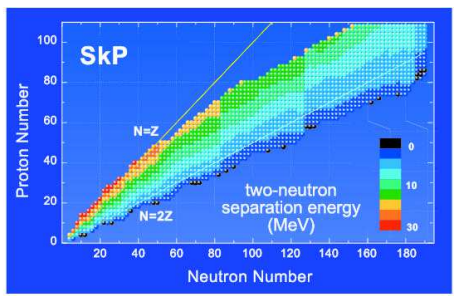}
\caption{(Color online) 
An example of a large-scale systematic density functional theory (DFT) 
calculation for complex nuclei: results of the deformed DFT calculations 
of two-neutron separation energies for 1553 particle-bound 
even-even nuclei with $Z\le 108$ and $N\le 188$ (from M. Stoitsov).
}
\end{center}
\label{fig4}
\end{figure}

\section{The next 10 years}

I believe that the next 10 years hold great promise for the physics of
nuclei and for moving toward the ultimate solution of the nuclear
problem: to obtain a comprehensive understanding of all nuclei and their
interactions.  On the experimental side, new facilities at RIKEN, GSI, 
and GANIL, and the future FRIB in the U.S. will make the discoveries that 
will challenge our theories and our understanding of many-body 
phenomena. 

From a theoretical point of view, I believe the key to 
reaching the goal is the pursuit of many-body theory and its 
connections to nuclear DFT.  We will see in the next few 
years the development of nuclear theory such
that robust and predictive ab initio calculations can be 
accomplished in medium-mass nuclei. The effort
to establish an improved density functional should allow for breakthrough
science in heavy nuclei and in nuclei 
along the various nucleosynthesis paths. The strides being
made to connect ab initio calculations in light- to medium-mass systems to 
nuclear DFT will enable us to finally rest the energy density functional on 
a firmer footing. The links from QCD to the underlying interaction will 
allow us to close the loop and reliably compute nuclei from the ground up. 

All of the efforts I described in the preceding section -- GFMC, NCSM, 
coupled-cluster efforts, and the quest to find the universal 
nuclear energy density functional -- require petascale computational 
capability. We have witnessed in the last five years tremendous 
growth in computing capability and many in our community continue to 
embrace these marvelous tools. The ingenuity of researchers
to utilize the largest of these machines undoubtedly 
will propel the field forward in our quest to understand nuclei. 

\label{}



\end{document}